\documentclass[conference]{IEEEtran}
\IEEEoverridecommandlockouts
\usepackage{orcidlink}
\usepackage[T1]{fontenc}
\usepackage{hyperref}
\usepackage{adjustbox}
\usepackage{multirow} 
\usepackage{float}
\usepackage{xfrac}
\usepackage{booktabs}
\usepackage{makecell}
\usepackage{array}
\usepackage{threeparttable}
\usepackage{cite}
\usepackage{amsmath,amssymb,amsfonts}
\usepackage{algorithmic}
\usepackage{graphicx}
\usepackage{textcomp}
\usepackage{xcolor}
\usepackage[utf8]{inputenc}
\DeclareUnicodeCharacter{1EF3}{\`y}
\usepackage{textcase} % Added to provide fallback for small caps
\def\BibTeX{{\rm B\kern-.05em{\sc i\kern-.025em b}\kern-.08em
    T\kern-.1667em\lower.7ex\hbox{E}\kern-.125emX}}

\usepackage{lipsum}
\usepackage{fancyhdr}
\fancypagestyle{sec}{\lhead{\tmpx}}

\fancypagestyle{arXiv}
{
   \fancyhf{}
   \chead{\textcolor{gray}{This paper has been accepted to be presented at the 2025 International Symposium on Circuits and Systems (ISCAS), \\London, UK, May 25–28, 2025}}
   \cfoot{ \fontsize{=9pt}{10pt}\selectfont  \textcolor{gray}{© 2025 IEEE. Personal use of this material is permitted. Permission from IEEE must be obtained for all other uses, in any current or future media, including reprinting/republishing this material for advertising or promotional purposes, creating new collective works, for resale or redistribution to servers or lists, or reuse of any copyrighted component of this work in other works}\\\fontsize{=10pt}{12pt}\selectfont\thepage}
   
}

\begin{document}

\title{DPD-NeuralEngine: A 22-nm 6.6-TOPS/W/mm$^2$ Recurrent Neural Network Accelerator for Wideband Power Amplifier Digital Pre-Distortion}

% \author{ 
% Anonymous Authors
% }

\author{ 
Ang~Li\textsuperscript{*}\orcidlink{0000-0003-3615-6755},
Haolin~Wu\textsuperscript{*},
Yizhuo~Wu\orcidlink{0009-0009-5087-7349},
Qinyu~Chen\orcidlink{0009-0005-9480-6164},
Leo~C.~N.~de~Vreede\orcidlink{0000-0002-5834-5461},
Chang~Gao\orcidlink{0000-0002-3284-4078}

\thanks{\textsuperscript{*}Ang Li and Haolin Wu contributed equally to this work.}
\thanks{Corresponding Author: Chang Gao (chang.gao@tudelft.nl)}
\thanks{A. Li, H. Wu, Y. Wu, L. C. N. de Vreede and C. Gao are with the Department of Microelectronics, Delft University of Technology, The Netherlands.}
\thanks{Q. Chen is with the Leiden Institute of Advanced Computer Science (LIACS), Leiden University, The Netherlands.}
}

\maketitle

\begin{abstract}
The increasing adoption of Deep Neural Network (DNN)-based Digital Pre-distortion (DPD) in modern communication systems necessitates efficient hardware implementations. This paper presents \texttt{DPD-NeuralEngine}, an ultra-fast, tiny-area, and power-efficient DPD accelerator based on a Gated Recurrent Unit (GRU) neural network (NN). Leveraging a co-designed software and hardware approach, our 22\,nm CMOS implementation operates at 2\,GHz, capable of processing I/Q signals up to 250\,MSps. Experimental results demonstrate a throughput of 256.5\,GOPS and power efficiency of 1.32\,TOPS/W with DPD linearization performance measured in Adjacent Channel Power Ratio (ACPR) of -45.3 dBc and Error Vector Magnitude (EVM) of -39.8 dB. To our knowledge, this work represents the first AI-based DPD application-specific integrated circuit (ASIC) accelerator, achieving a power-area efficiency (PAE) of 6.6\,TOPS/W/mm$^2$.
\end{abstract}

\begin{IEEEkeywords}
Deep Neural Network, Digital Pre-distortion, Software-Hardware Co-Design, ASIC, FPGA
\end{IEEEkeywords}

\section{Introduction}
\thispagestyle{arXiv}
The evolution of wireless communication systems towards higher data rates and broader bandwidths has intensified demands on transmitter digital backends (DBEs), potentially making them primary power consumers. As 5G and future 6G systems employ sophisticated modulation schemes and wider baseband bandwidths ($f_{BB}$), Digital Pre-Distortion (DPD) algorithms in DBEs must process data at sampling rates up to thousands of mega samples per second (MSps) to effectively linearize power amplifiers (PAs).

Massive Multiple-Input Multiple-Output (mMIMO) systems, enhancing spectral efficiency through numerous antennas, require efficient DBEs to handle increased computational loads~\cite{Richter2009,Richter2010}. However, power constraints in wireless systems, particularly in base stations and IoT devices, necessitate solutions with high power-area efficiency (PAE) measured in Operations per Second per Watt per square millimeter (OPS/W/mm\textsuperscript{2}).

Traditional DPD techniques, such as the generalized memory polynomial (GMP) model~\cite{Morgan2006}, struggle to meet linearization performance requirements for wideband PAs. Additionally, stringent frequency and latency constraints of advanced communication standards create a pressing need for dedicated DPD hardware accelerators that deliver high computational throughput while maintaining low power consumption and minimal silicon area.
\begin{figure}[t]
    \centering
    \includegraphics[width=0.9\linewidth]{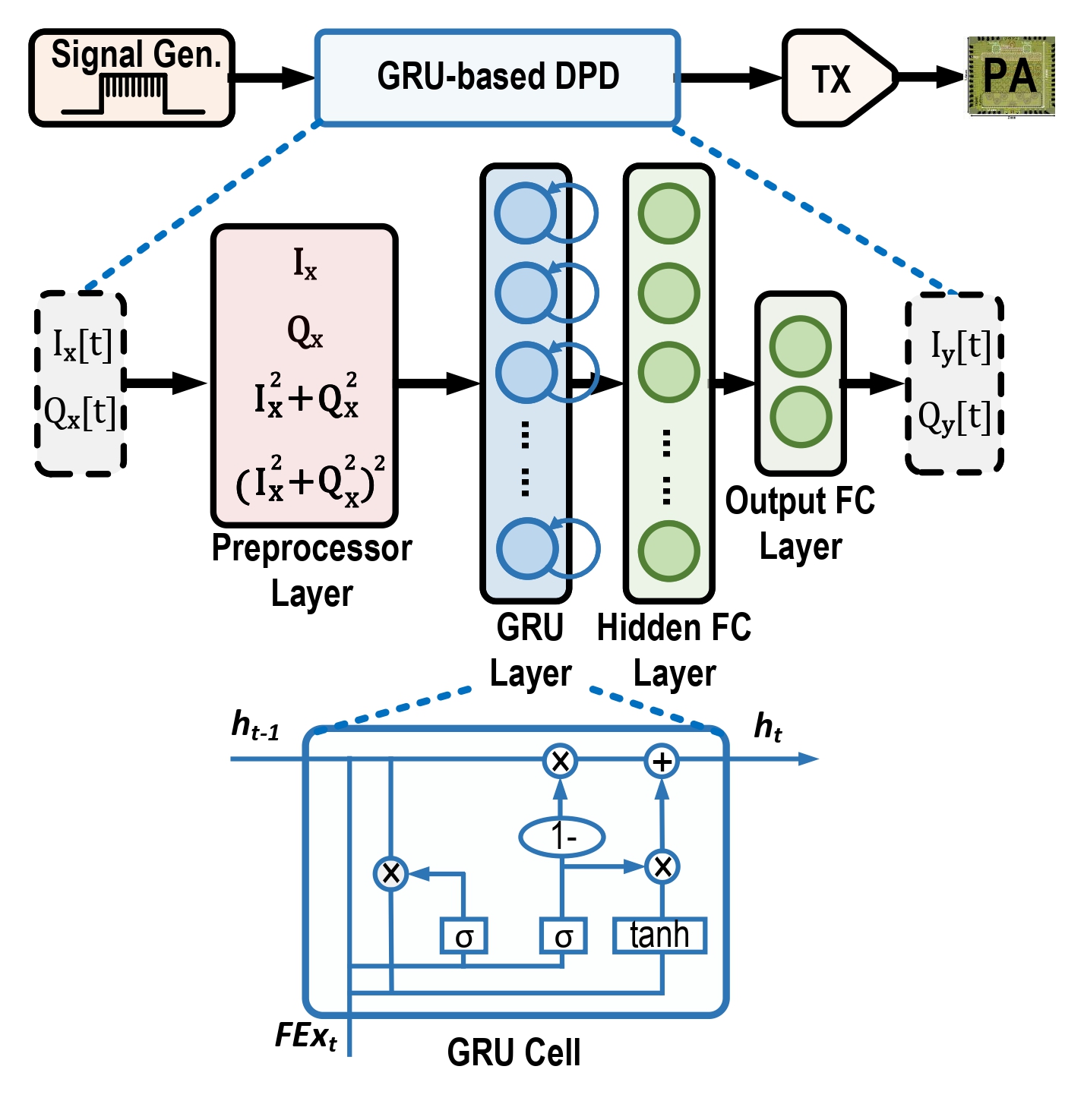}
    % \captionsetup{justification=centering}
    \caption{The GRU-RNN DPD architecture}
    \label{fig:structure}
\end{figure} 

Deep Neural Networks (DNNs) have shown promise in modeling complex PA nonlinearities for wideband systems. Early approaches like Time Delay Neural Networks (TDNNs)~\cite{Rawat2010,Zhang2019} paved the way for more sophisticated models such as VDLSTM and SVDLSTM~\cite{Li2020}, which leverage recurrent neural networks (RNNs) to capture PA dynamics. Recent developments include evaluation frameworks like OpenDPD~\cite{wu2024opendpd} and mixed-precision models such as MP-DPD~\cite{wu2024mp}.

However, hardware implementation of DNN-based DPD systems presents significant challenges. The computational and memory demands of DNNs impede real-time processing, particularly under the silicon area and power constraints of wireless SoCs~\cite{chuang2024role}. Current DPD FPGA implementations face challenges balancing power consumption and throughput when handling very high I/Q sample rates~\cite{Chen2006,Lin2011,Huang2019, Huang2020,Cappello2022,Yue2022,li2024gpu}. Similarly, prior FPGA-~\cite{chang2017iscas,han2017ese,gao2018deltarnn,li2019ernn,gao2020edgedrnn,gao2024spartus} and ASIC-based~\cite{kadetotad20208,shin201714,kim202223,lee2019,Han2016,Molendijk2023,Lin2020} DNN/RNN accelerators were designed for relatively low-bit-rate tasks, such as audio (hundreds of kbps) or video (hundreds of Mbps for 480p@30FPS), and are not optimized for DPD to process I/Q data streams with Gbps-level bit rates in wideband transmitters.

In this paper, we propose \texttt{DPD-NeuralEngine}, a 22nm RNN-based DPD Application-Specific Integrated Circuit (ASIC) accelerator achieving 6.6 TOPS/W/mm\textsuperscript{2} in running a Gated Recurrent Unit (GRU)-based DPD algorithm. The accelerator operates at 2.0\,GHz and can process signals with sampling rates up to 250\,MSps (3\,Gbps for 12-bit I/Q). To the best of our knowledge, this is the first AI-based DPD ASIC accelerator.
%-------------------------------------
\section{RNN-based DPD Algorithm}
%-------------------------------------
GRU-based RNNs can effectively model long-term dependencies in sequential data, making them ideal for DPD applications. As illustrated in \autoref{fig:structure}, our GRU-RNN DPD model comprises three layers: the preprocessor, GRU, and fully connected (FC) layers.

Initially, the input in-phase (\(I_x\)) and quadrature (\(Q_x\)) signals are processed to extract four features, forming the input feature vector \(\mathbf{x}_t\) at time \(t\):
\begin{equation}
\mathbf{x}_t = \begin{pmatrix}
I_{x,t} \\
Q_{x,t} \\
I_{x,t}^2 + Q_{x,t}^2 \\
\left( I_{x,t}^2 + Q_{x,t}^2 \right)^2
\end{pmatrix}
\end{equation}

These features are then input into the GRU layer, defined by the following equations:
\begin{align}
\mathbf{r}_t &= \sigma\left( \mathbf{W}_{ir} \mathbf{x}_t + \mathbf{b}_{ir} + \mathbf{W}_{hr} \mathbf{h}_{t-1} + \mathbf{b}_{hr} \right) \\
\mathbf{z}_t &= \sigma\left( \mathbf{W}_{iz} \mathbf{x}_t + \mathbf{b}_{iz} + \mathbf{W}_{hz} \mathbf{h}_{t-1} + \mathbf{b}_{hz} \right) \\
\mathbf{n}_t &= \tanh\left( \mathbf{W}_{in} \mathbf{x}_t + \mathbf{b}_{in} + \mathbf{r}_t \odot \left( \mathbf{W}_{hn} \mathbf{h}_{t-1} + \mathbf{b}_{hn} \right) \right) \\
\mathbf{h}_t &= (1 - \mathbf{z}_t) \odot \mathbf{n}_t + \mathbf{z}_t \odot \mathbf{h}_{t-1}
\end{align}

Here, \(\mathbf{x}_t\) represents the input feature vector at time \(t\), as defined in Equation (1), and \(\mathbf{h}_{t-1}\) is the previous hidden state. The reset gate \(\mathbf{r}_t\) and update gate \(\mathbf{z}_t\) control the flow of information, while \(\mathbf{n}_t\) generates the candidate hidden state.

The FC layer then maps the GRU's hidden state \(\mathbf{h}_t\) to the output predistorted signals \(I_y\) and \(Q_y\):
\begin{equation}
\begin{pmatrix}
I_{y,t} \\
Q_{y,t}
\end{pmatrix}
= \mathbf{W}_{\text{FC}} \mathbf{h}_t + \mathbf{b}_{\text{FC}}
\end{equation}

These outputs are subsequently converted to analog signals for amplification by the PA.
\begin{figure}[t!]
    \centering
    \includegraphics[width=0.48\textwidth]{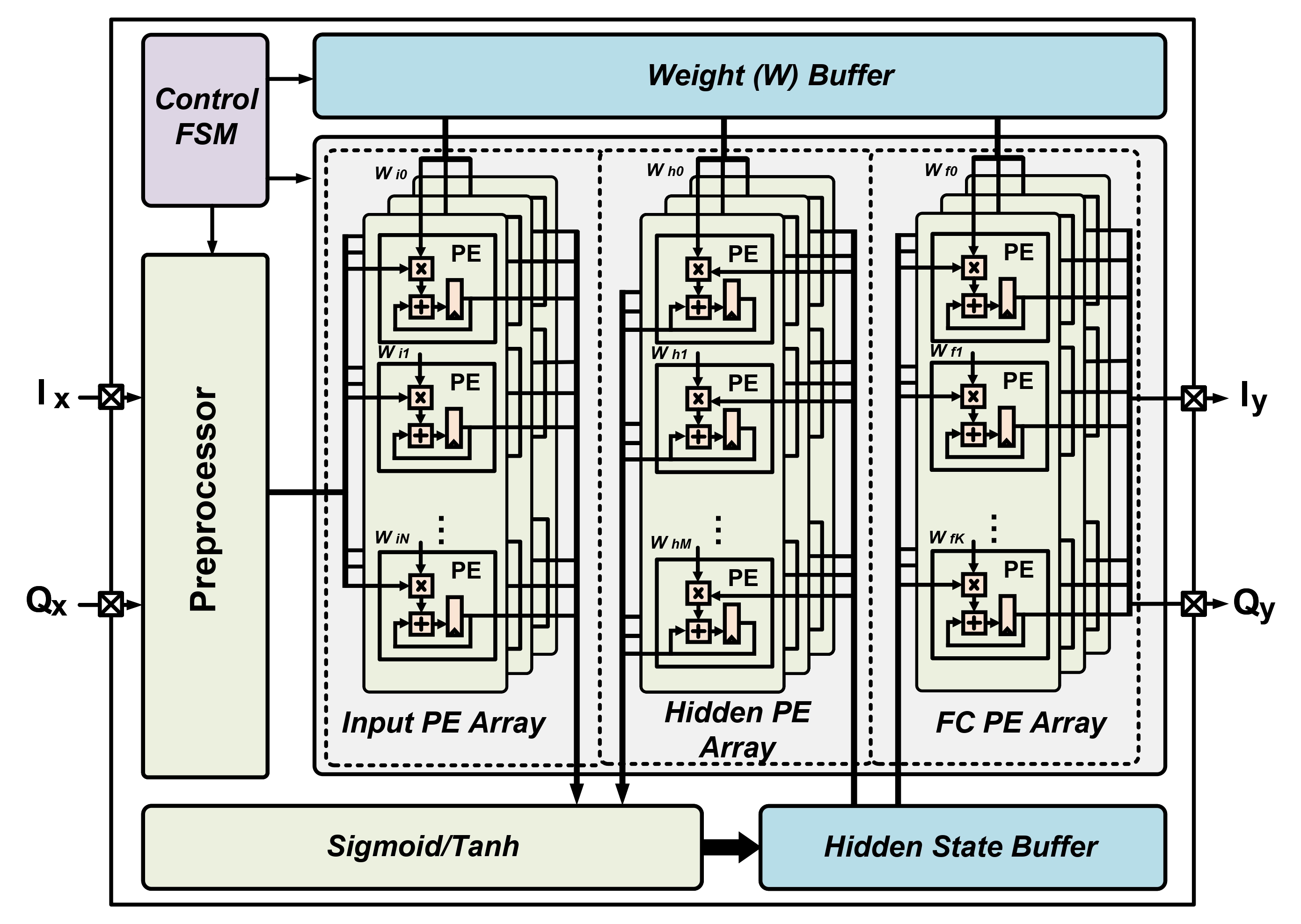}
    \caption{Microarchitecture of the GRU-RNN DPD hardware accelerator}
    \label{fig:Micro-architecture}
\end{figure}

%----------------------------------------
\section{DPD ASIC Accelerator Design}
%----------------------------------------
Building upon the GRU-RNN DPD model, we propose a hardware accelerator designed for real-time inference. The accelerator's microarchitecture comprises four primary components: a preprocessor, a Processing Element (PE) array, nonlinear function units, and memory buffers, all orchestrated by a central Finite State Machine (FSM).

\subsection{Microarchitecture}
The preprocessor uses 2 PEs to extract features from input I/Q signals, feeding them into the PE array, which consists of 156 PEs and is subdivided into input, hidden, and FC arrays, and performs matrix multiplications for the GRU and fully connected layers. Each PE executes Multiplication and Accumulation (MAC) operations, with varying levels of parallelism tailored to respective layer dimensions.

Nonlinear activation functions are implemented using efficient approximation methods, detailed in Section~\ref{sec:nonlinear}. The design incorporates two main buffers: a weight buffer storing fixed-point model parameters and a hidden state buffer for temporarily storing GRU computations.

\subsection{Nonlinear Function Approximation}\label{sec:nonlinear}

To address the computational complexity of sigmoid and tanh functions in hardware, we implement Piecewise Linear (PWL) approximations, namely \texttt{Hardsigmoid} and \texttt{Hardtanh}:
\begin{equation}
    \mathrm{Hardsigmoid}(x_i) = \left\{\begin{matrix}1,&x_i>2\\ \sfrac{x_i}{4}+\sfrac{1}{2}, &-2\leqslant x_i\leqslant 2\\ 0, &x_i<-2\end{matrix}\right.
\end{equation}
\begin{equation}
    \mathrm{Hardtanh}(x_i) = \left\{\begin{matrix}1,&x_i>1\\ x_i, &-1\leqslant x_i\leqslant 1\\ -1, &x_i<-1\end{matrix}\right.
\end{equation}
\noindent where $x_i$ is the $i$-th element of the input vector. This approach simplifies their hardware implementation to a series of comparators and shifters.

\subsection{Fixed-Point Data Representation}

To optimize hardware efficiency while maintaining computational accuracy, we employ a 12-bit Q2.10 fixed-point data format (2 integer bits and 10 fractional bits) for both NN weights and activations and also the input and output I/Q data. The GRU-RNN DPD model is trained using Quantization-Aware Training (QAT) to minimize accuracy loss compared to the original floating-point model, which will be further discussed in Section~\ref{sec:model_accuracy}.
\begin{figure}[t]
    \centering
    \includegraphics[width=0.9\linewidth]{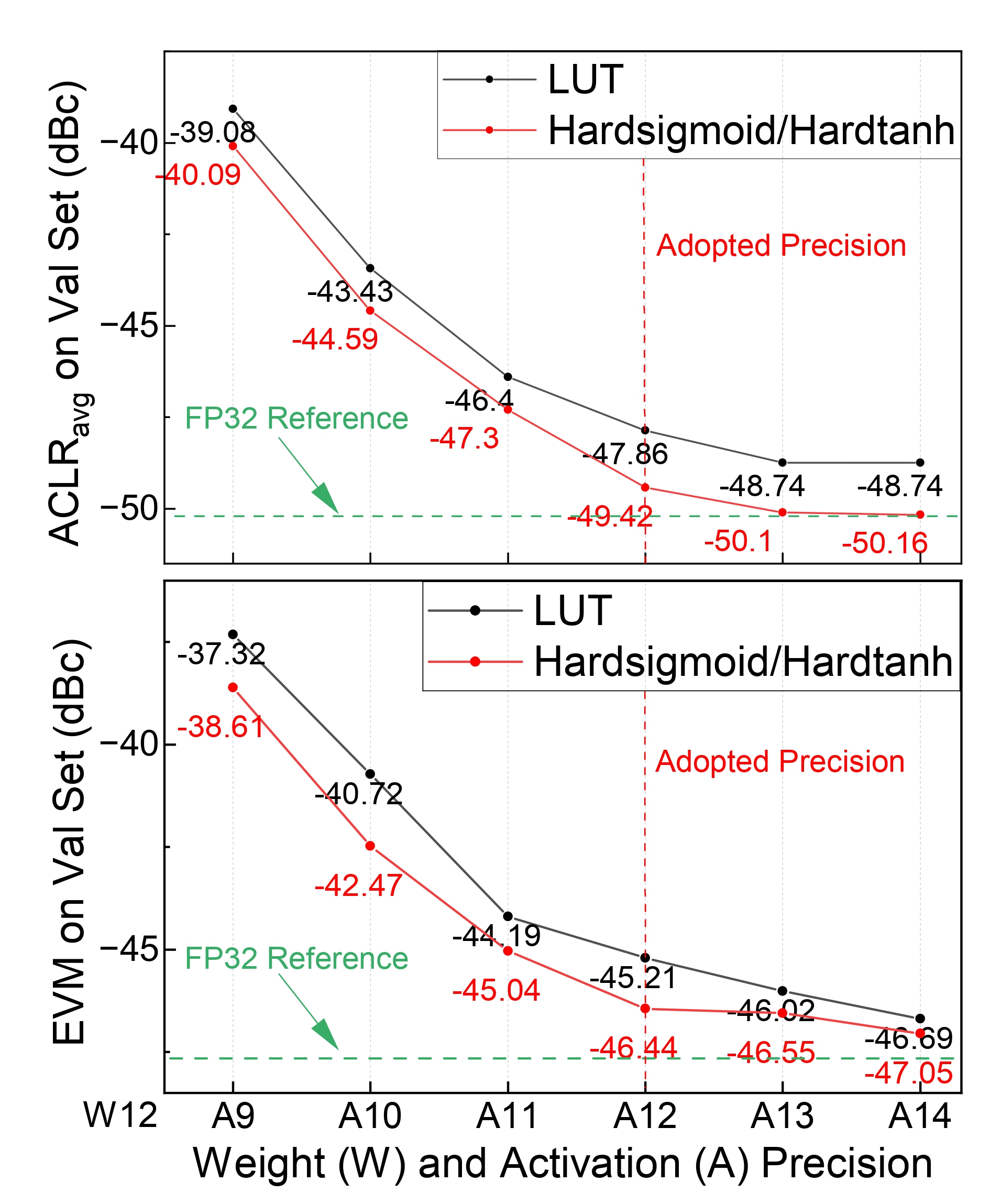}
    % \captionsetup{justification=centering}
    \caption{Comparison of GRU-RNN DPD performance between using LUT-based and \texttt{Hardsigmoid}/\texttt{Hardtanh} activation functions vs. precisions in \texttt{OpenDPD}~\cite{wu2024opendpd} simulations.}
    \label{fig:compare software}
\end{figure}
The subsequent section will detail our experimental setup and results, demonstrating the efficacy of its performance in wideband PA linearization.
%------------------------------------
\section{Experimental Results}
%------------------------------------
\subsection{Experimental Setup}
\subsubsection{Software}
The GRU-RNN DPD model, with 4 input features, 10 hidden units, and 1 hidden layer (502 parameters total), is evaluated using the 200 MHz OpenDPD dataset~\cite{wu2024opendpd} and a new 80\,MHz, 64-QAM, OFDM signal dataset (8.2dB PAPR). Both datasets use a 60-20-20 train-validation-test split. The latter dataset trains the model used in real measurements with a Keysight M8190A generator, linear pre-amplifier, GaN Doherty PA, and R\&S-FSW43 analyzer. The GaN Doherty's average output power is 40\,dBm before and after DPD.

The training utilizes an NVIDIA RTX 2050 Laptop GPU. QAT runs for 300 epochs using \(ReduceLROnPlateau\) scheduler (initial lr=1$\times$10\textsuperscript{-3}), with batch size 64, frame length 50, and stride 1.

\subsubsection{Hardware}
The design is first simulated on a Digilent PYNQ Board (Zynq-7020 FPGA-SoC) for verification and resource estimation. It is then implemented as an ASIC using GlobalFoundries 22FDX-PLUS FD-SOI technology. Cadence tools are used: Genus for synthesis, Innovus for placement and routing, and Xcelium for simulations. Performance and power results are derived from switching-activity-annotated post-layout simulations.

\begin{table}[t]
\caption{Utilization of DPD-NeuralEngine FPGA Emulation}
\label{tab:hardware_utilization}
\resizebox{\columnwidth}{!}{%
\begin{tabular}{c|cccc}
\Xhline{3\arrayrulewidth}
 &
  \begin{tabular}[c]{@{}c@{}}LUT\end{tabular} &
  \begin{tabular}[c]{@{}c@{}}FF\end{tabular} &
  DSP &
  BRAM \\ \hline
Available &
  53200 &
  106400 &
  220 &
  140 \\ \hline
\begin{tabular}[c]{@{}c@{}}Used \\ (LUT-Sig./Tanh)\end{tabular} &
  \begin{tabular}[c]{@{}c@{}}20522 \\ (38.7\%)\end{tabular} &
  \begin{tabular}[c]{@{}c@{}}3969 \\ (3.7\%)\end{tabular} &
  \begin{tabular}[c]{@{}c@{}}85 \\ (38.6\%)\end{tabular} &
  \begin{tabular}[c]{@{}c@{}}0 \\ (0\%)\end{tabular} \\ \hline
\begin{tabular}[c]{@{}c@{}}Used \\ (Hard-Sig./Tanh)\end{tabular} &
  \begin{tabular}[c]{@{}c@{}}5439 \\ (10.2\%)\end{tabular} &
  \begin{tabular}[c]{@{}c@{}}3156 \\ (3.0\%)\end{tabular} &
  \begin{tabular}[c]{@{}c@{}}95 \\ (43.2\%)\end{tabular} &
  \begin{tabular}[c]{@{}c@{}}0 \\ (0\%)\end{tabular} \\
\Xhline{3\arrayrulewidth}
\end{tabular}
}
\end{table}
\subsection{Results and Evaluation}
\subsubsection{Model Accuracy}
\label{sec:model_accuracy}
\autoref{fig:compare software} shows a comparison of model accuracy between using Look-Up Table (LUT)-based activation functions and using \texttt{Hardsigmoid}/\texttt{Hardtanh} functions at different precision levels, with the 32-bit floating-point model as the reference baseline. The figure indicates that, at the same weight and activation precision, the GRU-DPD model with \texttt{Hardsigmoid}/\texttt{Hardtanh} functions trained by QAT can achieve higher linearization performance than the LUT method, with an ACPR/EVM improvement of 1-2\,dB. A precision sweep for quantized models reveals that quantizing weights and activations to 12 bits provides an optimal balance between accuracy and hardware overhead.
%---------------------------------
% Comparison Table
%---------------------------------
% \input{tabs/tab_compare.tex}
\begin{table*}[t]
\begin{minipage}{\textwidth}
\caption{Comparison with the State-of-the-Art DPD Hardware Accelerators and measured signal quality.}
\renewcommand{\arraystretch}{1.8}  % Increase row spacing
\fontsize{14}{16}\selectfont  % Increase font size to 12pt with 14pt line spacing
\label{tab:compare}
\resizebox{\textwidth}{!}{%
\begin{tabular}{c|cccccccccccc|ccc}
\Xhline{6\arrayrulewidth}
 &
  \multicolumn{12}{c|}{\textbf{DPD Hardware Specification and Performance}} &
  \multicolumn{3}{c}{\textbf{Signal Config. and Quality\textsuperscript{1}}} \\ \hline
 &
  \begin{tabular}[c]{@{}c@{}}Architecture\\ ~\end{tabular} &
  \begin{tabular}[c]{@{}c@{}}Tech.\\ (nm)\end{tabular} &
  \begin{tabular}[c]{@{}c@{}}Model\\ ~\end{tabular} &
  \begin{tabular}[c]{@{}c@{}}Precision\textsuperscript{2}\\(bits) ~\end{tabular} &
  \begin{tabular}[c]{@{}c@{}}\#Param\\ ~\end{tabular} &
  \begin{tabular}[c]{@{}c@{}} OP/S\textsuperscript{2}\\ ~\end{tabular} &
  \begin{tabular}[c]{@{}c@{}}$\mathbf{f_{clk}}$\\ (MHz)\end{tabular} &
  \begin{tabular}[c]{@{}c@{}}$\mathbf{f_{s, I/Q}}$\\ (MSps)\end{tabular} &
  \begin{tabular}[c]{@{}c@{}}Latency\\ (ns)\end{tabular} &
  \begin{tabular}[c]{@{}c@{}}Throughput\textsuperscript{3}\\ (GOPS)\end{tabular} &
  \begin{tabular}[c]{@{}c@{}}Power\textsuperscript{4}\\ (W)\end{tabular} &
  \begin{tabular}[c]{@{}c@{}}Efficiency\textsuperscript{3}\\ (GOPS/W)\end{tabular} &
  \begin{tabular}[c]{@{}c@{}}$\mathbf{f_{BB}}$\\ (MHz)\end{tabular} &
  \begin{tabular}[c]{@{}c@{}}ACPR \\ (dBc)\end{tabular} &
  \begin{tabular}[c]{@{}c@{}}EVM \\ (dB)\end{tabular} \\ \hline
This Work &
  ASIC &
  22 &
  RNN &
  W12A12 &
  502 &
  1,026 &
  2,000 &
  250 &
  \textbf{7.5} &
  256.5 &
  \textbf{0.20} &
  \textbf{1,315.4} &
  60 &
  -45.3 &
  -39.8 \\
~\cite{Huang2020} &
  FPGA (UltraScale+) &
  16 &
  GMP &
  W?A16 &
  36 &
  17 &
  300 &
  \textbf{2,400} &
  - &
  $\sim$40.8 &
  0.96 &
  $\sim$42.5 &
  400 &
  -44.7 &
  -39.2 \\
~\cite{Cappello2022} &
  FPGA (Zynq-7000) &
  28 &
  MP &
  W?A16 &
  9 &
  30 &
  250 &
  250 &
  40 &
  $\sim$7.5 &
  0.23 &
  $\sim$32.6 &
  20 &
  -49.0 &
  - \\
~\cite{Yue2022} &
  FPGA (Virtex-7) &
  28 &
  GMP &
  W?A16 &
  38 &
  149 &
  - &
  400 &
  - &
  $\sim$59.6 &
  0.89 &
  $\sim$67.0 &
  100 &
  -46.45 &
  - \\
~\cite{li2024gpu} &
  GPU (RTX 4080) &
  \textbf{5} &
  TDNN &
  FP32 &
  909 &
  $\sim$1,818 &
  $\sim$2,300 &
  1,000 &
  - &
  \textbf{$\sim$1,818} &
  $\leq$320 &
  $\geq$5.7 &
  200 &
  -45.2 &
  -35.34 \\
\Xhline{6\arrayrulewidth}
\end{tabular}
}

\vspace{1ex}
\footnotesize{\textsuperscript{1} Signal quality of PA outputs after applying DPD. Note absolute values here are incomparable due to different signal configurations and PA hardware.} \\
\footnotesize{\textsuperscript{2} Precision of Weights (W) and Activation (A). “?” indicates values are not reported.} \\
\footnotesize{\textsuperscript{3} OP/S denotes Operations Per I/Q Sample (floating-point or fixed-point). OPS (Giga Operations Per Second) is calculated as GOPS = OP/S $\times~f_{s, I/Q}$.} \\
\footnotesize{\textsuperscript{4} Reported total on-chip power (dynamic + static). For~\cite{li2024gpu}, Thermal Design Power (TDP)~\cite{rtx4080} is used here as measured power is not reported.} \\
% \footnotesize{\textsuperscript{4}~\cite{Huang2020} utilizes a parallel architecture with a parallelism factor of 8, resulting in the required operations per sample to \#OP/8.}

\end{minipage}
\vspace{-8pt}

\end{table*}
\begin{figure}[t]
    \centering
    \includegraphics[width=0.9\linewidth]{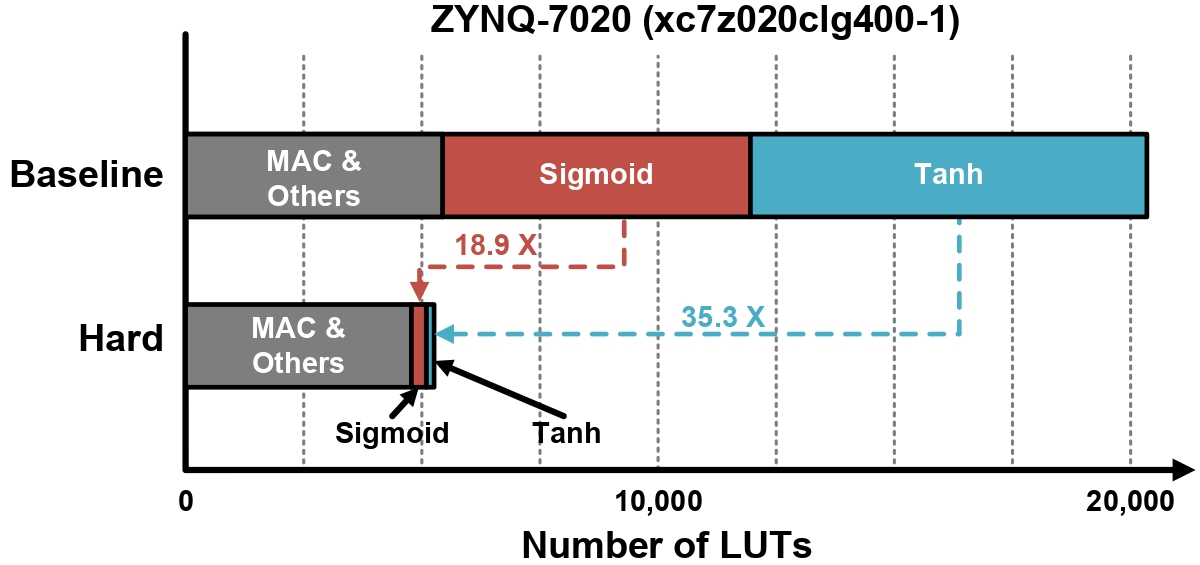}
    \caption{Breakdown of LUT Usage on ZYNQ-7020: Baseline (LUT-Sigmoid/Tanh) vs. Hard (Hard-Sigmoid/Tanh).}
    \label{fig:fpga_lut}
\end{figure}

\subsection{FPGA Emulation}
\autoref{tab:hardware_utilization} shows the resource utilization of FPGA-emulated \texttt{DPD-NeuralEngine} using a baseline LUT-based and \texttt{Hardsigmoid}/\texttt{Hardtanh} activation functions. \autoref{fig:fpga_lut} shows that LUT-based activation function implementations consume even more FPGA-LUTs (over 20,000) than PEs for MAC operations. In contrast, \texttt{Hardsigmoid}/\texttt{Hardtanh} functions significantly reduce FPGA-LUT usage for both \texttt{sigmoid} and \texttt{tanh} by 18.9$\times$ and 35.3$\times$, respectively, reducing the total FPGA-LUT usage to around 5,500, demonstrating a significant reduction of their area cost.
\begin{table}[t]
\caption{Comparison with Prior RNN/DNN ASICs}
\label{tab:compare_asic}
\resizebox{\columnwidth}{!}{%
\begin{tabular}{c|ccccccc|c}
\Xhline{3\arrayrulewidth}
 &
  \cite{kadetotad20208} &
  \cite{shin201714} &
  \cite{kim202223} &
  \cite{lee2019} &
  \cite{Han2016} &
  \cite{Molendijk2023} &
  \cite{Lin2020} &
  \begin{tabular}[c]{@{}c@{}}This \\ work\end{tabular} \\ \hline
\begin{tabular}[c]{@{}c@{}}Technology\\  (nm)\end{tabular} &
  65 &
  65 &
  65 &
  65 &
  45 &
  22 &
  \textbf{7} &
  22 \\ \hline
\begin{tabular}[c]{@{}c@{}}$\mathrm{f}_{clk}$\\  (MHz)\end{tabular} &
  80&
  200&
  0.25 &
  200 &
  800 &
  300 &
  880 &
  \textbf{2,000}\\ \hline
\begin{tabular}[c]{@{}c@{}}Weight Prec.\\  (bits)\end{tabular} &
  32 &
  32 &
  32 &
  16 &
  4 &
  8 &
  8 &
  12 \\ \hline
\begin{tabular}[c]{@{}c@{}}Area \\ (mm\textsuperscript{2})\end{tabular} &
  7.7 &
  16.0 &
  0.4 &
  16 &
  40.8 &
  3.0 &
  3.0 &
  \textbf{0.2} \\ \hline
\begin{tabular}[c]{@{}c@{}}Supply \\ (V)\end{tabular} &
  1.1 &
  1.1 &
  0.75 &
  1.1 &
  - &
  0.5 &
  0.575 &
  0.9 \\ \hline
\begin{tabular}[c]{@{}c@{}}Power \\ (mW)\end{tabular} &
  67 &
  21 &
  \textbf{0.02} &
  297 &
  590 &
  31 &
  174 &
  195 \\ \hline
\begin{tabular}[c]{@{}c@{}}Throughput \\ (GOPS)\end{tabular} &
  165 &
  25 &
  0.004 &
  346 &
  102 &
  77 &
  \textbf{3,604} &
  257 \\ \hline
\begin{tabular}[c]{@{}c@{}}Power Efficiency \\ (TOPS/W)\end{tabular} &
  2.45 &
  1.19 &
  0.17 &
  3.08 &
  0.17 &
  2.47 &
  \textbf{6.83} &
  1.32 \\ \hline
\begin{tabular}[c]{@{}c@{}}Area Efficiency \\ (GOPS/mm²)\end{tabular} &
  21.3 &
  1.6 &
  0.01 &
  21.6 &
  2.5 &
  25.8 &
  1,185.7 &
  \textbf{1,282.5} \\ \hline
\begin{tabular}[c]{@{}c@{}}PAE \\ (TOPS/W/mm²)\end{tabular} &
  0.32 &
  0.07 &
  0.40 &
  0.07 &
  0.004 &
  0.83 &
  2.25 &
  \textbf{6.58} \\ 
\Xhline{3\arrayrulewidth}
\end{tabular}
}
\end{table}

\begin{figure}[t]
    \centering
    \includegraphics[width=0.86\linewidth]{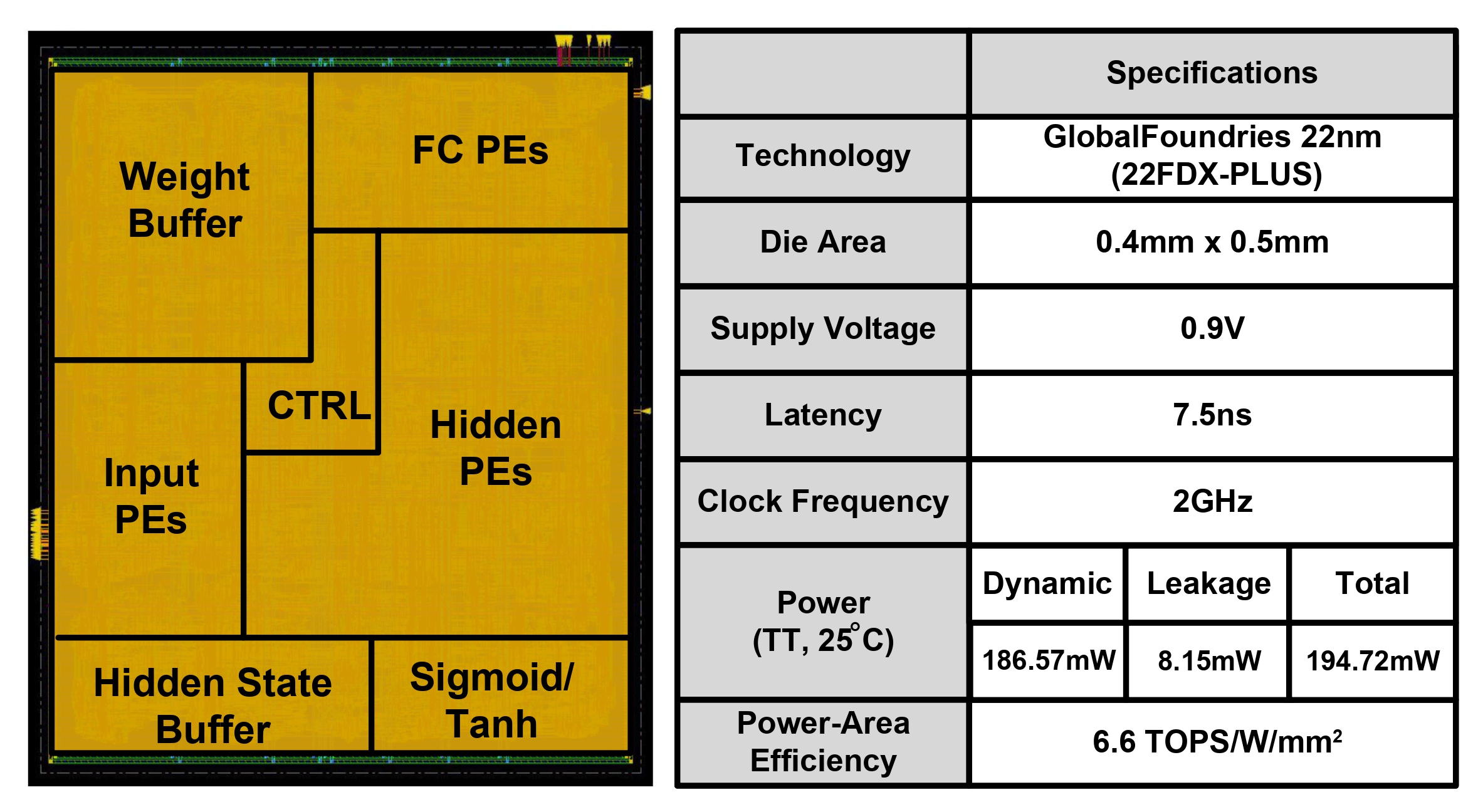}
    \caption{Post-layout specification of DPD-NeuralEngine.}
    \label{fig:layout}
\end{figure}

\subsection{ASIC Implementation}
\autoref{fig:layout} summarizes the post-layout area (0.2\,mm\textsuperscript{2}) and performance of \texttt{DPD-NeuralEngine} operating at a clock frequency ($f_{clk}$) of 2\,GHz with a supply voltage of 0.9\,V. With a total power consumption of 195\,mW, the design achieves a latency of 7.5\,ns and 256.5\,GOPS throughput, capable of handling real-time DPD with an I/Q sample rate of 250\,MSps; thus achieving a PAE of 6.6 TOPS/W/mm\textsuperscript{2}.

\subsection{Comparison With Previous Work}
~\autoref{tab:compare} shows a comparison between the proposed accelerator and other state-of-the-art DPD hardware designs. Most previous works utilized FPGAs and memory polynomial (MP)-based DPD models, with only one neural network DPD implementation on a GPU~\cite{li2024gpu}.
Our proposed \texttt{DPD-NeuralEngine} ASIC achieves the lowest on-chip power consumption while achieving the fastest latency and the highest power efficiency. Although the GPU-based approach~\cite{li2024gpu} offers superior throughput at 1,818 GOPS, it costs significantly higher power consumption due to the unnecessary redundancy of a desktop RTX 4080 GPU ($\leq$320 W). Furthermore, this work exhibits competitive signal quality metrics, with an ACPR of -45.3 dBc and EVM of -39.8 dB at a baseband frequency of 60 MHz. 

We also compare \texttt{DPD-NeuralEngine} to classic prior RNN/DNN ASICs as shown in~\autoref{tab:compare_asic}. Our design achieves the highest PAE over others, which is important since DPD has stringent area and power consumption requirements simultaneously. This is aided by the compact model, which allows unnecessary flexibility to be sacrificed to co-design specialized hardware, thereby boosting PAE. The closest design is a 7\,nm DNN accelerator~\cite{Lin2020}; though achieving higher power efficiency thanks to more advanced technology, lower bit precision, and larger chip scale, they have worse PAE due to the unnecessary programmability for ultra-high-speed DPD application with tight area budget.

The reported results demonstrate the potential of NN-based DPD ASIC accelerators to balance performance, power efficiency, and signal quality, making them ideal for next-generation wireless communication systems.

%--------------------------
\section{Conclusion}
%--------------------------
This paper presents an efficient ASIC implementation of a GRU-RNN DPD accelerator for wideband power amplifier linearization. The reported efficiency numbers significantly outperform existing FPGA- and GPU-based DPD implementations. As we approach 6G, integrating advanced AI algorithms with efficient hardware is crucial. Our ASIC-based approach demonstrates the potential of neural network-based DPD accelerators to optimize performance, power efficiency, and signal quality for future wireless communication systems.

%--------------------------
\section*{Acknowledgement}
%--------------------------
We thank GlobalFoundries for providing the 22FDX PDK and Ampleon Netherlands B.V. for providing the GaN PA.

\bibliographystyle{IEEEtran}
\bibliography{refs.bib}

\begin{thebibliography}{10}
\providecommand{\url}[1]{#1}
\csname url@samestyle\endcsname
\providecommand{\newblock}{\relax}
\providecommand{\bibinfo}[2]{#2}
\providecommand{\BIBentrySTDinterwordspacing}{\spaceskip=0pt\relax}
\providecommand{\BIBentryALTinterwordstretchfactor}{4}
\providecommand{\BIBentryALTinterwordspacing}{\spaceskip=\fontdimen2\font plus
\BIBentryALTinterwordstretchfactor\fontdimen3\font minus \fontdimen4\font\relax}
\providecommand{\BIBforeignlanguage}[2]{{%
\expandafter\ifx\csname l@#1\endcsname\relax
\typeout{** WARNING: IEEEtran.bst: No hyphenation pattern has been}%
\typeout{** loaded for the language `#1'. Using the pattern for}%
\typeout{** the default language instead.}%
\else
\language=\csname l@#1\endcsname
\fi
#2}}
\providecommand{\BIBdecl}{\relax}
\BIBdecl

\bibitem{Richter2009}
F.~Richter, A.~J. Fehske, and G.~P. Fettweis, ``Energy efficiency aspects of base station deployment strategies for cellular networks,'' in \emph{Proceedings of the 70th IEEE Vehicular Technology Conference Fall (VTC 2009-Fall)}, 2009, pp. 1--5.

\bibitem{Richter2010}
F.~Richter and G.~Fettweis, ``Cellular mobile network densification utilizing micro base stations,'' in \emph{Proceedings of the 2010 IEEE International Conference on Communications (ICC 2010)}, 2010, pp. 1--6.

\bibitem{Morgan2006}
D.~R. Morgan, Z.~Ma, J.~Kim, M.~G. Zierdt, and J.~Pastalan, ``A generalized memory polynomial model for digital predistortion of rf power amplifiers,'' \emph{IEEE Transactions on Signal Processing}, vol.~54, no.~10, pp. 3852--3860, Oct 2006.

\bibitem{Rawat2010}
M.~Rawat, K.~Rawat, and F.~M. Ghannouchi, ``Adaptive digital predistortion of wireless power amplifiers/transmitters using dynamic real-valued focused time-delay line neural networks,'' \emph{IEEE Transactions on Microwave Theory and Techniques}, vol.~58, no.~1, pp. 95--104, Jan 2010.

\bibitem{Zhang2019}
Y.~Zhang, Y.~Li, F.~Liu, and A.~Zhu, ``Vector decomposition based time-delay neural network behavioral model for digital predistortion of rf power amplifiers,'' \emph{IEEE Access}, vol.~7, pp. 91\,559--91\,568, 2019.

\bibitem{Li2020}
H.~Li, Y.~Zhang, G.~Li, and F.~Liu, ``Vector decomposed long short-term memory model for behavioral modeling and digital predistortion for wideband rf power amplifiers,'' \emph{IEEE Access}, vol.~8, pp. 63\,780--63\,789, 2020.

\bibitem{wu2024opendpd}
Y.~Wu, G.~D. Singh, M.~Beikmirza, L.~C. De~Vreede, M.~Alavi, and C.~Gao, ``Opendpd: An open-source end-to-end learning \& benchmarking framework for wideband power amplifier modeling and digital pre-distortion,'' in \emph{2024 IEEE International Symposium on Circuits and Systems (ISCAS)}.\hskip 1em plus 0.5em minus 0.4em\relax IEEE, 2024, pp. 1--5.

\bibitem{wu2024mp}
Y.~Wu, A.~Li, M.~Beikmirza, G.~D. Singh, Q.~Chen, L.~C. de~Vreede, M.~Alavi, and C.~Gao, ``Mp-dpd: Low-complexity mixed-precision neural networks for energy-efficient digital predistortion of wideband power amplifiers,'' \emph{IEEE Microwave and Wireless Technology Letters}, 2024.

\bibitem{chuang2024role}
\BIBentryALTinterwordspacing
K.~Chuang, ``Role of {AI/ML} in {PA} linearization for next {G} wireless,'' Keynote abstract, IEEE International Microwave Symposium (IMS), 2024, accessed: Oct. 8, 2024. [Online]. Available: \url{https://ims-ieee.org/sites/ims2019/files/content_images/ims2024_keynote_abstract_kevinchuang_adi_v4.pdf}
\BIBentrySTDinterwordspacing

\bibitem{Chen2006}
H.~H. Chen, C.~H. Lin, P.~C. Huang, and J.~T. Chen, ``Joint polynomial and look-up-table predistortion power amplifier linearization,'' \emph{IEEE Transactions on Circuits and Systems II: Express Briefs}, vol.~53, no.~8, pp. 612--616, Aug 2006.

\bibitem{Lin2011}
H.~Lin, X.~Guo, Z.~Zhang, J.~Cao, F.~Lin, D.~Wei, and Y.~Fang, ``Optimization design of fpga-based look-up-tables for linearizing rf power amplifiers,'' in \emph{2011 International Conference on Electronics, Communications and Control (ICECC)}, Ningbo, China, 2011, pp. 2731--2734.

\bibitem{Huang2019}
H.~Huang, J.~Xia, and S.~Boumaiza, ``Parallel-processing-based digital predistortion architecture and fpga implementation for wide-band 5g transmitters,'' in \emph{2019 IEEE MTT-S International Microwave Conference on Hardware and Systems for 5G and Beyond (IMC-5G)}, Atlanta, GA, USA, 2019, pp. 1--3.

\bibitem{Huang2020}
------, ``Novel parallel-processing-based hardware implementation of baseband digital predistorters for linearizing wideband 5g transmitters,'' \emph{IEEE Transactions on Microwave Theory and Techniques}, vol.~68, no.~9, pp. 4066--4076, 2020.

\bibitem{Cappello2022}
T.~Cappello, G.~Jindal, J.~Nunez-Yanez, and K.~Morris, ``Power consumption and linearization performance of a bit- and frequency-scalable am/am am/pm pre-distortion on fpga,'' in \emph{2022 International Workshop on Integrated Nonlinear Microwave and Millimetre-Wave Circuits (INMMiC)}, Cardiff, United Kingdom, 2022, pp. 1--3.

\bibitem{Yue2022}
Y.~Li, X.~Wang, and A.~Zhu, ``Reducing power consumption of digital predistortion for rf power amplifiers using real-time model switching,'' \emph{IEEE Transactions on Microwave Theory and Techniques}, vol.~70, no.~3, pp. 1500--1508, 2022.

\bibitem{li2024gpu}
\BIBentryALTinterwordspacing
W.~Li, R.~Criado, W.~Thompson, K.~Chuang, G.~Montoro, and P.~L. Gilabert, ``Gpu-based implementation of pruned artificial neural networks for digital predistortion linearization of wideband power amplfiers,'' Jul. 2024. [Online]. Available: \url{http://dx.doi.org/10.36227/techrxiv.172227112.20453024/v1}
\BIBentrySTDinterwordspacing

\bibitem{chang2017iscas}
A.~X.~M. Chang and E.~Culurciello, ``Hardware accelerators for recurrent neural networks on fpga,'' in \emph{2017 IEEE International Symposium on Circuits and Systems (ISCAS)}, 2017, pp. 1--4.

\bibitem{han2017ese}
S.~Han, J.~Kang, H.~Mao, Y.~Hu, X.~Li, Y.~Li, D.~Xie, H.~Luo, S.~Yao, Y.~Wang \emph{et~al.}, ``Ese: Efficient speech recognition engine with sparse lstm on fpga,'' in \emph{Proceedings of the 2017 ACM/SIGDA International Symposium on Field-Programmable Gate Arrays}, 2017, pp. 75--84.

\bibitem{gao2018deltarnn}
C.~Gao \emph{et~al.}, ``Deltarnn: A power-efficient recurrent neural network accelerator,'' in \emph{Proceedings of the 2018 ACM/SIGDA International Symposium on Field-Programmable Gate Arrays}, 2018, pp. 21--30.

\bibitem{li2019ernn}
Z.~Li, C.~Ding, S.~Wang, W.~Wen, Y.~Zhuo, C.~Liu, Q.~Qiu, W.~Xu, X.~Lin, X.~Qian, and Y.~Wang, ``E-rnn: Design optimization for efficient recurrent neural networks in fpgas,'' in \emph{2019 IEEE International Symposium on High Performance Computer Architecture (HPCA)}, 2019, pp. 69--80.

\bibitem{gao2020edgedrnn}
C.~Gao, A.~Rios-Navarro, X.~Chen, S.-C. Liu, and T.~Delbruck, ``Edgedrnn: Recurrent neural network accelerator for edge inference,'' \emph{IEEE Journal on Emerging and Selected Topics in Circuits and Systems}, vol.~10, no.~4, pp. 419--432, 2020.

\bibitem{gao2024spartus}
C.~Gao, T.~Delbruck, and S.-C. Liu, ``Spartus: A 9.4 top/s fpga-based lstm accelerator exploiting spatio-temporal sparsity,'' \emph{IEEE Transactions on Neural Networks and Learning Systems}, vol.~35, no.~1, pp. 1098--1112, 2024.

\bibitem{kadetotad20208}
D.~Kadetotad, S.~Yin, V.~Berisha, C.~Chakrabarti, and J.-s. Seo, ``An 8.93 tops/w lstm recurrent neural network accelerator featuring hierarchical coarse-grain sparsity for on-device speech recognition,'' \emph{IEEE Journal of Solid-State Circuits}, vol.~55, no.~7, pp. 1877--1887, 2020.

\bibitem{shin201714}
D.~Shin, J.~Lee, J.~Lee, and H.-J. Yoo, ``14.2 dnpu: An 8.1 tops/w reconfigurable cnn-rnn processor for general-purpose deep neural networks,'' in \emph{2017 IEEE International Solid-State Circuits Conference (ISSCC)}.\hskip 1em plus 0.5em minus 0.4em\relax IEEE, 2017, pp. 240--241.

\bibitem{kim202223}
K.~Kim, C.~Gao, R.~Graça, I.~Kiselev, H.-J. Yoo, T.~Delbruck, and S.-C. Liu, ``{A 23µW Solar-Powered Keyword-Spotting ASIC with Ring-Oscillator-Based Time-Domain Feature Extraction},'' in \emph{IEEE International Solid-State Circuits Conference (ISSCC)}, Feb. 2022, pp. 370--371.

\bibitem{lee2019}
J.~Lee, C.~Kim, S.~Kang, D.~Shin, S.~Kim, and H.-J. Yoo, ``Unpu: An energy-efficient deep neural network accelerator with fully variable weight bit precision,'' \emph{IEEE Journal of Solid-State Circuits}, vol.~54, no.~1, pp. 173--185, 2019.

\bibitem{Han2016}
S.~Han, X.~Liu, H.~Mao, J.~Pu, A.~Pedram, M.~A. Horowitz, and W.~J. Dally, ``Eie: efficient inference engine on compressed deep neural network,'' in \emph{Proceedings of the 43rd International Symposium on Computer Architecture}, ser. ISCA '16.\hskip 1em plus 0.5em minus 0.4em\relax IEEE Press, 2016, p. 243–254.

\bibitem{Molendijk2023}
M.~J. Molendijk, F.~A.~M. de~Putter, M.~D. Gomony, P.~Jääskeläinen, and H.~Corporaal, ``Braintta: A 28.6 tops/w compiler programmable transport-triggered nn soc,'' in \emph{2023 IEEE 41st International Conference on Computer Design (ICCD)}, 2023, pp. 78--85.

\bibitem{Lin2020}
C.-H. Lin, C.-C. Cheng, Y.-M. Tsai, S.-J. Hung, Y.-T. Kuo, P.~H. Wang, P.-K. Tsung, J.-Y. Hsu, W.-C. Lai, C.-H. Liu, S.-Y. Wang, C.-H. Kuo, C.-Y. Chang, M.-H. Lee, T.-Y. Lin, and C.-C. Chen, ``7.1 a 3.4-to-13.3tops/w 3.6tops dual-core deep-learning accelerator for versatile ai applications in 7nm 5g smartphone soc,'' in \emph{2020 IEEE International Solid-State Circuits Conference - (ISSCC)}, 2020, pp. 134--136.

\bibitem{rtx4080}
\BIBentryALTinterwordspacing
Nvidia, ``Nvidia geforce rtx 4080 super en rtx 4080 grafische kaarten.'' [Online]. Available: \url{https://www.nvidia.com/nl-nl/geforce/graphics-cards/40-series/rtx-4080-family/}
\BIBentrySTDinterwordspacing

\end{thebibliography}

\end{document}